\documentclass[fleqn,twoside]{article}
\usepackage{espcrc2}

\usepackage{graphicx}
\usepackage[figuresright]{rotating}

\newcommand{\AmS}{{\protect\the\textfont2
  A\kern-.1667em\lower.5ex\hbox{M}\kern-.125emS}}
\newcommand{\BR}[1]{\ensuremath{{\rm BR}(#1)}}
\newcommand{\SN}[2]{\ensuremath{#1\times10^{#2}}}

\newcommand{\Fig}[1]{Fig.~\ref{#1}}

\newcommand{\Tab}[1]{Table~\ref{#1}}

\newcommand{\etal}{et al.}

\title{Precision tests of the Standard Model  with leptonic and semileptonic kaon decays}

\author{B. Sciascia~\address[LNF]{Laboratori Nazionali di Frascati dell'INFN} 
on behalf of the FlaviaNet Kaon Working 
Group~\thanks{WWW access at www.lnf.infn.it/wg/vus/; for sake of 
completeness and brevity for all references we refer to the Note written 
by the FlaviaNet Kaon WG~\cite{Flavia2008}.}}
       
\begin{document}

\begin{abstract}
This paper presents the analysis of leptonic and semileptonic kaon decays data
done by the FlaviaNet Kaon Working group, as described 
in~\cite{Flavia2008}.
Data include all recent results by BNL-E865, KLOE, KTeV, ISTRA+, and NA48.
Experimental results are
critically reviewed and combined, taking into account theoretical
(both analytical and numerical) constraints on the semileptonic
kaon form factors. 
We report on a very accurate determination of $V_{us}$ as well as
on many other tests of the SM which 
can be performed with leptonic and semileptonic $K$ decays.
\end{abstract}

\maketitle

\section{Introduction}
In the Standard Model, SM, transition rates of semileptonic processes such as 
$d^i \to  u^j \ell \nu$, with $d^i$ ($u^j$) being a generic 
down (up) quark, can be computed with high accuracy in terms 
of the Fermi coupling $G_F$ and the elements $V_{ji}$ of the 
Cabibbo-Kobayashi Maskawa (CKM) matrix. 
Measurements of the transition rates provide therefore
precise determinations of the fundamental SM couplings. 

A detailed analysis of semileptonic decays offers also 
the possibility to set stringent constraints on  new physics scenarios.
While within the SM all $d^i \to  u^j \ell \nu$ transitions 
are ruled by the same CKM coupling $V_{ji}$ (satisfying 
the unitarity condition $\sum_k |V_{ik}|^2 =1$) and 
$G_F$ is the same coupling appearing in the muon decay, 
this is not necessarily true beyond the SM. 
Setting bounds on the violations of CKM unitarity, 
violations of lepton universality, and deviations from 
the $V-A$ structure, allows us to put significant 
constraints on various new-physics scenarios
(or eventually find evidences of new physics). 

In the case of leptonic and semileptonic $K$ decays these tests 
are particularly significant given the large amount of
data recently collected by several experiments: 
BNL-E865, KLOE, KTeV, ISTRA+, and NA48. 
The analysis of these data provides precise determination of fundamental SM couplings,
sets stringent SM test almost free from hadronic uncertainties, and
finally can discriminate between new physics scenarios.
The high statistical precision of measurements  and the detailed information
on kinematical distributions have pushed a substantial progress on the theory side,
in particular the theoretical error on hadronic form factors has been reduced
at the 1\% level.

The paper is organized as follows. First in Sec.~\ref{BRfits}
we present fits to world data on the leading branching ratios and lifetimes,
for $K_L$, $K_S$, and $K^\pm$ mesons. Sec.~\ref{slopes} summarizes
the status of the knowledge of form factor slopes from $K_{\ell 3}$ decays.
The physics results obtained are described in Sec.~\ref{resulta}, in particular
the measurement of $|V_{us}f_+(0)|$.
Finally, to the special role
of $\Gamma(K_{e2}^\pm)/\Gamma(K_{\mu 2}^\pm)$ ratio is devoted the Sec.~\ref{ke2}.

\section{Experimental data: BRs and lifetime}
\label{BRfits}
Numerous measurements of the principal kaon BRs, or of various ratios
of these BRs, have been published recently. For the purposes of evaluating
$|V_{us}f_+(0)|$, these data can be used in a PDG-like fit to the BRs and lifetime,
so all such measurements are interesting.
A detailed description to
the fit procedure and the references of all experimental input used 
can be found in Ref.~\cite{Flavia2008}.

For $K_L$ the results are given in table~\ref{tab:KLBR}, while 
table~\ref{tab:KpmBR} gives the results for $K^\pm$.
\begin{table}
\begin{center}
\begin{tabular}{l|c|r}
Parameter & Value & $S$ \\
\hline
\BR{K_{e3}} & 0.4056(7) & 1.1 \\
\BR{K_{\mu3}} & 0.2705(7) & 1.1 \\
\BR{3\pi^0} & 0.1951(9) & 1.2 \\
\BR{\pi^+\pi^-\pi^0} & 0.1254(6) & 1.1 \\
\BR{\pi^+\pi^-} & \SN{1.997(7)}{-3} & 1.1 \\
\BR{2\pi^0} & \SN{8.64(4)}{-4} & 1.3 \\
\BR{\gamma\gamma} & \SN{5.47(4)}{-4} & 1.1 \\
$\tau_L$ & 51.17(20)~ns & 1.1 \\
\end{tabular}
\end{center}
\vskip 0.3cm
\caption{\label{tab:KLBR}
Results of fit to $K_L$ BRs and lifetime.}
\end{table}
For the $K_S$, the fit is dominated by the KLOE measurements of $BR(K_S\to\pi e\nu)$ and
of $BR(\pi^+\pi^-)/BR(\pi^0\pi^0)$. These, together with
the constraint that the $K_S$ BRs must add to unity, and the assumption of
universal lepton couplings, completely determine the $K_S$ leading BRs
In particular, $\BR{K_S\to\pi e\nu} = \SN{7.046(91)}{-4}$.
For $\tau_{K_S}$ we use \SN{0.8958}{-10}~s, where this is the non-$CPT$
constrained fit value from the PDG.
%
\begin{table}
\begin{center}
\begin{tabular}{l|c|r}
Parameter & Value & $S$ \\
\hline
\BR{K_{\mu2}}      & 63.57(11)\%   & 1.1 \\
\BR{\pi\pi^0}      & 20.64(8)\%   & 1.1 \\
\BR{\pi\pi\pi}     &  5.595(31)\%  & 1.0 \\
\BR{K_{e3}}        &  5.078(26)\%    & 1.2 \\
\BR{K_{\mu3}}      &  3.365(27)\%  & 1.7 \\
\BR{\pi\pi^0\pi^0} &  1.750(26)\%  & 1.1 \\
$\tau_\pm$         & 12.384(19)~ns & 1.7 \\
\end{tabular}
\end{center}
\vskip 0.3cm
\caption{\label{tab:KpmBR}
Results of fit to $K^\pm$ BRs and lifetime.}
\end{table}

\section{Experimental data: $K_{\ell 3 }$ form factors}
\label{slopes}
The hadronic  $K \to \pi$ matrix element of the vector current
is described by two form factors (FFs),  $f_+(t)$ and  $f_0(t)$.
By construction,  $f_0(0)=f_+(0)$.
In order to compute the phase space integrals 
we need experimental or theoretical inputs about the  $t$-dependence of FF.
In principle,  Chiral Perturbation Theory (ChPT) 
and Lattice QCD are useful tools to set theoretical constraints.
However, in practice the  $t$-dependence of the FFs at present 
is better determined by measurements and by combining measurements 
and dispersion relations. 
Many approaches have been used, and all have been described in detail in~\cite{Flavia2008}.
Here we list only the averages of quadratic fit results for $K_{e3}$ and $K_{\mu3}$ 
slopes (\Tab{tab:l3ff}) used to determine $|V_{us}|f_+(0)$.
\begin{table}
\begin{center}
\begin{tabular}{l|c}
\hline\hline
                                 & $K_L$ and $K^-$ \\
\hline
Measurements                     & 16		   \\
$\chi^2/{\rm ndf}$               & 54/13 $(7\times 10^{-7})$ \\
$\lambda_+'\times 10^3 $         & $24.9\pm1.1$ ($S=1.4$) \\
$\lambda_+'' \times 10^3 $       & $1.6\pm0.5$ ($S=1.3$)  \\
$\lambda_0\times 10^3 $          & $13.4\pm1.2$ ($S=1.9$) \\
$\rho(\lambda_+',\lambda_+'')$   & $-0.94$                \\
$\rho(\lambda_+',\lambda_0)$     & $+0.33$                \\
$\rho(\lambda_+'',\lambda_0)$    & $-0.44$                \\
$I(K^0_{e3})$                    & 0.15457(29)	      \\
$I(K^\pm_{e3})$                  & 0.15892(30)	      \\
$I(K^0_{\mu3})$                  & 0.10212(31)	      \\
$I(K^\pm_{\mu3})$                & 0.10507(32)	      \\
$\rho(I_{e3},I_{\mu3})$          & $+0.63$             \\
\hline\hline
\end{tabular}
\end{center}
\caption{Averages of quadratic fit results for $K_{e3}$ and $K_{\mu3}$ slopes.}
\label{tab:l3ff}
\end{table}

\section{Physics results}
\label{resulta}
\subsection{Determination of $| V_{us}|f_{+}(0)$ and
 $| V_{us}|/| V_{ud}|f_K/f_\pi$}
The value of $|V_{us}|f_{+}(0)$ has been determined from 
the decay rate of kaon semileptonic decays (see~\cite{Flavia2008} for the detailed
decomposition).
using the world average values reported in previous sections
for lifetimes, branching ratios and phase space integrals.
\begin{figure}[t]
\centering
\includegraphics[width=0.7\linewidth]{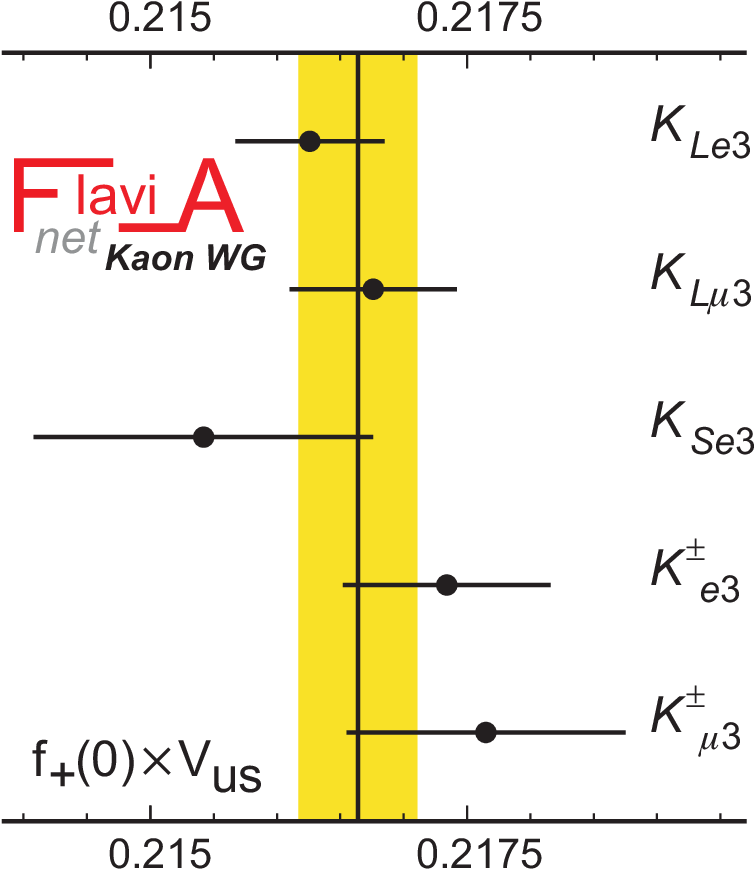}
\caption{ Display of $|V_{us}|f_{+}(0) $ for all channels. 
\label{fig:Vusf0} }
\end{figure}
The results  are shown in figure~\ref{fig:Vusf0} 
for  $K_L\to\pi e\nu$, $K_L\to\pi\mu\nu$,
$K_S\to\pi e\nu$, $K^\pm\to\pi e\nu$, $K^\pm\to\pi\mu\nu$, 
and for the combination.
The average,   
$|V_{us}|f_+(0)=0.21664(48)$, has an uncertainty of about of $0.2\%$.  
The results from the five modes are in good agreement, the
fit probability is 58\%. 
In particular, comparing the values of $|V_{us}|f_{+}(0)$
obtained from $K^0_{\ell3}$ and $K^\pm_{\ell3}$ we obtain
a value of the SU(2) breaking correction 
$\delta^K_{SU(2)_{exp.}}=2.9(4)\%$
in agreement with the CHPT calculation $\delta^K_{SU(2)}= 2.36(22)\%$.
Moreover, recent analyzes
on the so-called violations of Dashen's theorem in the kaon 
electromagnetic mass splitting point to $\delta^{K}_{SU(2)}$ 
values of about $3\%$.

The test of  Lepton Flavor Universality (LFU) between  
$K_{e3}$ and $K_{\mu3}$ modes constraints a possible
anomalous lepton-flavor dependence in the leading 
weak vector current. It can therefore be compared
to similar tests in $\tau$ decays, but is different from the 
LFU tests in the helicity-suppressed modes $\pi_{l2}$ and $K_{l2}$.
The results on the parameter
$r_{\mu e} = R_{K_{\mu3}/K_{e3}}^{\rm{Exp}}/R_{K_{\mu3}/K_{e3}}^{\rm{SM}}$ is
$r_{\mu e} = 1.0043 \pm 0.0052$,
in excellent agreement with lepton universality. 
With a precision of $0.5\%$ the test in $K_{l3}$ decays 
has now reached the sensitivity of other determinations:
$r_{\mu e}(\tau) = 1.0005 \pm 0.0041$ and
$r_{\mu e}(\pi) = 1.0042 \pm 0.0033$~\cite{PDG06}

An independent  determination of $V_{us}$ is obtained from $K_{\ell2}$ decays. 
The most important mode is  $K^+\to\mu^+\nu$, which has been recently 
updated by KLOE reaching a relative uncertainty of about $0.3\%$.  
Hadronic uncertainties are minimized considering the ratio 
$\Gamma(K^+\to\mu^+\nu)/\Gamma(\pi^+\to\mu^+\nu)$.
Using the world average values
of BR($K^\pm\to\mu^\pm\nu$) and of $\tau^\pm$ given in Section~\ref{BRfits}
and the value of $\Gamma(\pi^\pm\to\mu^\pm\nu)=38.408(7)~\mu s^{-1}$
from~\cite{PDG06} we obtain:
$|V_{us}|/|V_{ud}|f_K/f_\pi = 0.2760 \pm  0.0006$.

\subsection{Theoretical estimates of $f_+(0)$ and $f_K/f_\pi$ }
The main obstacle in transforming these highly precise determinations of 
$|V_{us}|f_{+}(0)$ and 
$|V_{us}|/|V_{ud}|f_K/f_\pi$ into a determination of 
$|V_{us}|$ at the per-mil level are the theoretical 
uncertainties on the hadronic parameters $f_+(0)$ and $f_K/f_\pi$.
This hadronic quantity cannot be computed in perturbative QCD, but
it is highly constrained by $SU(3)$ and chiral symmetry. 
In the chiral limit and, more generally, in the $SU(3)$ limit
($m_u=m_d=m_s$) the conservation of the vector current
implies $f_+(0)$=1. Expanding around the chiral limit in powers 
of light quark masses we can write
$f_+(0)= 1 + f_2 + f_4 + \ldots$
where $f_2$ and $f_4$ are the NLO and 
NNLO corrections in ChPT.  The Ademollo--Gatto theorem implies that
$(f_+(0)-1)$ is at least of second order in the breaking of $SU(3)$ 
This in turn implies 
that  $f_2$ is free from the uncertainties  of the $\mathcal{O}(p^4)$ counterterms in ChPT, 
and it  can be computed with high accuracy: $f_2=-0.023$. 
The difficulties in 
estimating $f_+(0)$ begin with $f_4$ or at $\mathcal{O}(p^6)$ in the chiral expansion.
Several analytical approaches to determine $f_4$
have been attempted over the years, 
essentially confirming the original estimate by Leutwyler and Roos.
The benefit of these new results, obtained using 
more sophisticated techniques, lies in the fact that a 
better control over the systematic uncertainties of the calculation
has been obtained. However, the size of the error is still around or 
above $1\%$, which is not comparable  to the $0.2\%$
accuracy which has been reached for $|V_{us}|f_+(0)$.

Recent progress in lattice QCD gives us more optimism in the reduction of 
the error on $f_+(0)$ below the $1\%$ level. 
Most of the currently available 
lattice QCD  results have been obtained with relatively heavy pions and  
the chiral extrapolation represents the dominant source of uncertainty.
There is a general trend of 
 lattice QCD results to be slightly lower than  analytical approaches. 
An important step in the reduction of the error associated to the 
chiral extrapolation  has been recently made by 
the UKQCD-RBC collaboration. 
Their preliminary  result $f_+(0)=0.964(5)$
is obtained from the unquenched study  with 
$N_F=2+1$ flavors, with an action that has good chiral properties 
on the lattice even at finite lattice 
spacing (domain-wall quarks). They also reached pions masses ($\geq 330$ MeV) 
much lighter than that used in  previous studies of $f_+(0)$. The 
overall error is estimated to be  $~0.5\%$, which is very encouraging.

In contrast to the semileptonic vector form factor, the pseudoscalar
decay constants are not protected by the Ademollo--Gatto theorem and 
receive corrections linear in the quark masses. Expanding 
$f_K/f_\pi$ in power of quark masses, in analogy to $f_+(0)$, 
$f_K/f_\pi= 1 + r_2 + \ldots$
one finds that the $\mathcal{O}(p^4)$ contribution $r_2$ is 
already affected by local contributions and cannot be unambiguously 
predicted in ChPT. As a result, in the determination of $f_K/f_\pi$
lattice QCD 
has essentially no competition from purely analytical approaches. 
The  present overall accuracy is about $1\%$. 
The  novelty are the new  lattice results with 
$N_F=2+1$ dynamical quarks  and  pions as light as  $280$~MeV, 
obtained by using the so-called staggered quarks.
These analyzes cover a broad range of lattice spacings (i.e.~$a$=0.06 and 0.15 fm) and  
is performed on sufficiently large physical volumes ($m_\pi L\geq 5.0$). 
It should be stressed, however, that the sensitivity of 
$f_K/f_\pi$ to lighter pions is larger 
than in the computation of $f_+(0)$ and that  chiral extrapolations are far more 
demanding in this case.
In the following analysis we will use as reference value the MILC-HPQCD
result $f_K/f_\pi=1.189(7)$. 

\subsection{Test of CKM unitarity}
\begin{figure}[t]
\centering
\includegraphics[width=0.7\linewidth]{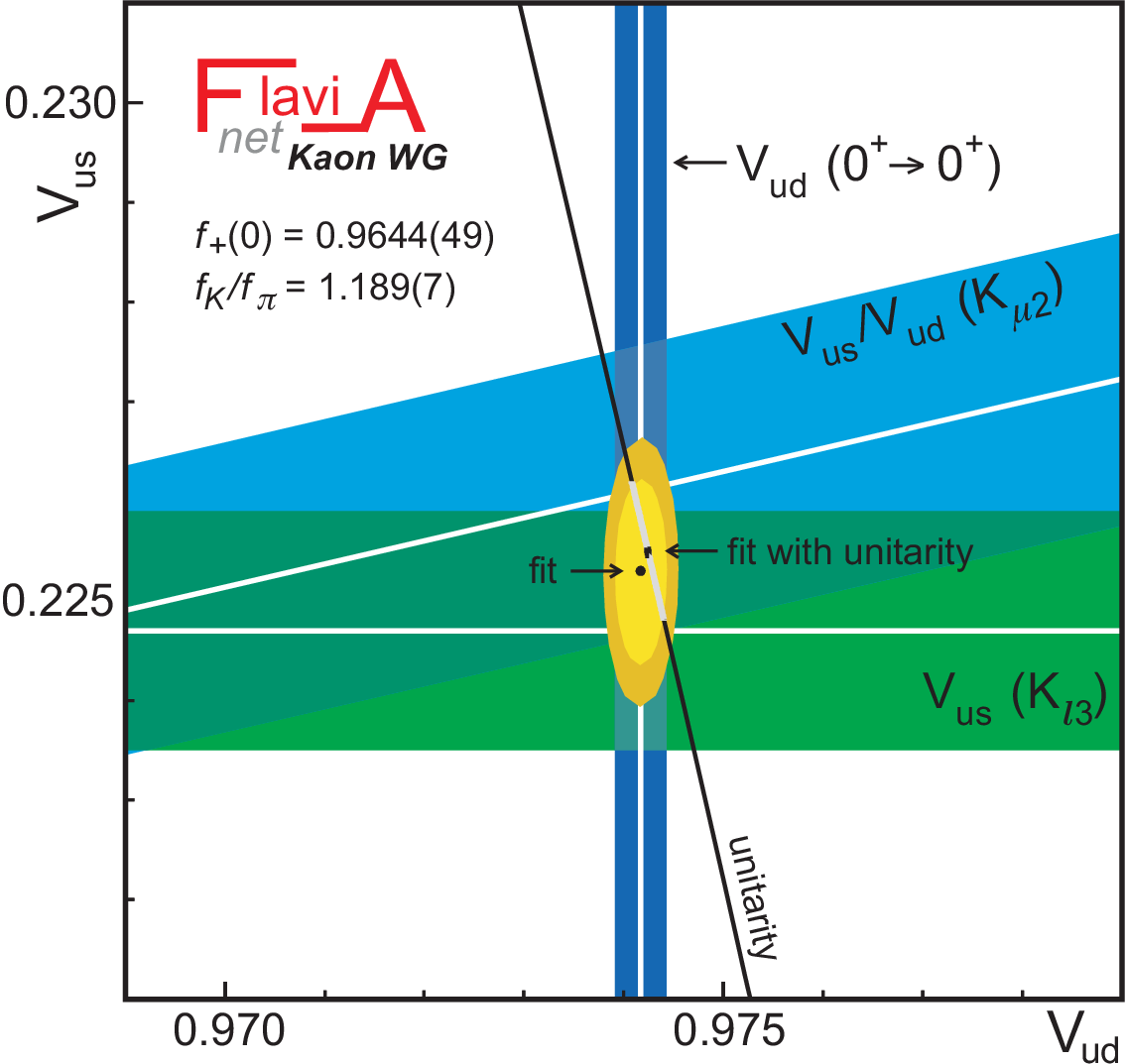}
\caption{\label{fig:vusuni} Results of fits to $|V_{ud}|$, $|V_{ux}|$, and $|V_{us}|/|V_{ud}|$.}
\end{figure}

To determine $|V_{us}|$ and   $|V_{ud}|$
 we use the value $|V_{us}| f_{+}(0)=0.2166(5)$ ,
the result $|V_{us}|/|V_{ud}|f_K/f_\pi = 0.2760(6)$, 
$f_+(0) = 0.964(5)$, and $f_K/f_\pi = 1.189(7)$. 
From the above we find:
$|V_{us}|=  0.2246\pm  0.0012$ from $K_{\ell 3}$ only, and
$|V_{us}|/|V_{ud}|= 0.2321\pm  0.0015$ from $K_{\ell 2}$ only.
These determinations can be used in a fit together with the 
the recent evaluation of $V_{ud}$ from
$0^+\to0^+$ nuclear beta decays: $|V_{ud}|$=0.97418$\pm$0.00026. 
This global fit gives $V_{ud} = 0.97417(26)$ and $V_{us} = 0.2253(9)$,
with $\chi^2/{\rm ndf} = 0.65/1$ (42\%). This result does not make use 
of CKM unitarity. If the  unitarity constraint is included, 
the fit gives $V_{us}=0.2255(7)$ and $\chi^2/{\rm ndf}=0.80/2$ (67\%).
Both results are illustrated in \Fig{fig:vusuni}.
The test of CKM unitarity can be also interpreted as a test of universality of
the lepton and quark gauge couplings.
Using the results of the fit (without imposing unitarity) we obtain:
$G_{\rm CKM} \equiv G_\mu \left[ |V_{ud}|^2+|V_{us}|^2+|V_{ub}|^2 \right]^{1/2}
= (1.1662 \pm  0.0004)\times 10^{-5}\  {\rm GeV}^{-2}$,
in perfect agreement with the value obtained from the measurement
of the muon lifetime:
$G_{\mu} = (1.166371 \pm  0.000007)\times 10^{-5}\  {\rm GeV}^{-2}.$
The current accuracy of the lepton-quark universality
sets important constraints on model building beyond the SM.
For example, the presence of  a $Z^\prime$ would affect the relation between 
$G_{\rm CKM}$ and $G_{\mu}$. In case of a $Z^\prime$ from $SO(10)$ grand unification theories
we obtain  $m_{Z^\prime}>700$~GeV at 95\% CL, to be compared with the  $m_{Z^\prime}>720$~GeV
bound  set through the direct collider searches~\cite{PDG06}.
In a similar way, the unitarity constraint also provides useful bounds in various 
supersymmetry-breaking scenarios. 

\subsection{$K_{\ell 2}$ sensitivity to new physics}
\begin{figure}[t]
\centering
\includegraphics[width=0.7\linewidth]{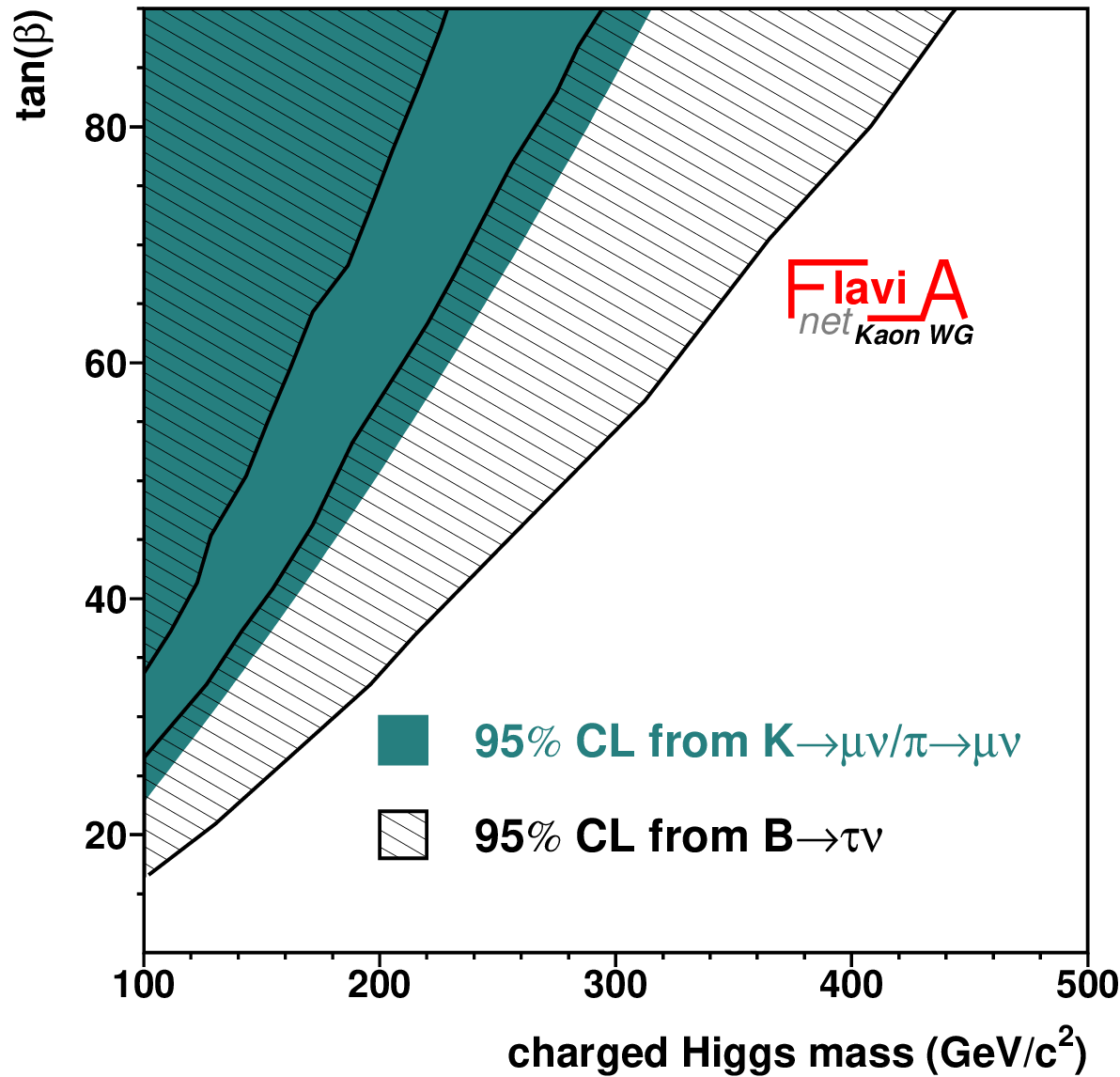}
\caption{\label{fig:higgskmunu}
Excluded region in the charged Higgs mass-$\tan\beta$ plane.
The region excluded by  $B\to \tau \nu $ is also indicated.}
\end{figure}

A particularly interesting test is the comparison of the $|V_{us}|$ 
value extracted from the helicity-suppressed $K_{\ell 2}$ decays
with respect to the value extracted from the  helicity-allowed $K_{\ell 3}$  modes.
To reduce theoretical uncertainties from $f_K$ and electromagnetic 
corrections in $K_{\ell 2}$, we exploit the ratio $BR(K_{\ell2})/BR(\pi_{\ell2})$ and 
we study the quantity
$$
R_{l23}=\left|\frac{V_{us}(K_{\ell 2})}{V_{us}(K_{\ell 3})}
\frac{V_{ud}(0^+\to 0^+)}{V_{ud}(\pi_{\ell 2})}\right|\,.
$$
Within the SM, $R_{l23}=1$, while deviation from 1 can be induced by 
non-vanishing scalar- or  right-handed currents.
Notice that in $R_{l23}$ the  hadronic uncertainties enter through  $(f_K/f_\pi)/f_+(0)$.
In the case of effect of scalar currents due to a charged Higgs, 
the unitarity relation between 
$|V_{ud}|$ extracted from $0^+\to0^+$ nuclear beta decays and $|V_{us}|$ extracted from 
$K_{\ell3}$ remains valid as soon as form factors are experimentally determined.
This constrain  together with the experimental information of $\log C^{MSSM}$ 
can be used in the global fit to improve the accuracy of the determination 
of $R_{l23}$, which in this scenario turns to be 
$\left. R_{l23} \right|^{\rm exp}_{\rm scalar} =  1.004 \pm  0.007$.
Here $(f_K/f_\pi)/f_+(0)$ has been fixed from lattice. This ratio
is the key quantity to be improved in order to reduce 
present uncertainty on $R_{l23}$. 
This  measurement of $R_{l23}$ can be used to set bounds
on the charged Higgs mass and $\tan\beta$.
Figure \ref{fig:higgskmunu} shows the excluded region at 95\%
CL in the $M_H$--$\tan\beta$ plane.
The measurement of  BR($B \to \tau \nu$) 
can be also used to set a similar  bound in the  $M_H$--$\tan\beta$ plane. 
While $B\to\tau \nu$ can exclude quite an extensive region of this plane,
there is an uncovered region in the exclusion corresponding 
to a destructive interference between the charged-Higgs 
and the SM amplitude. This region is fully covered by the $K\to \mu \nu$ result.

\subsection{A test of lattice calculation}
\label{sec:CTtest}

\begin{figure}[t]
\centering
\includegraphics[width=0.7\linewidth]{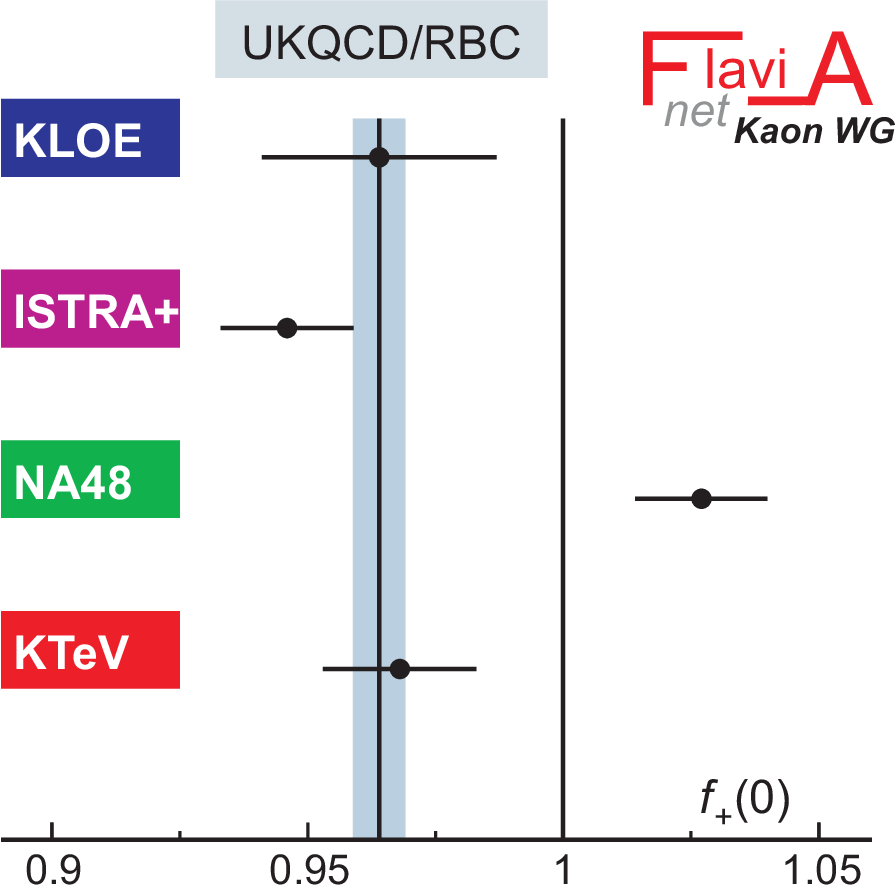}
\caption{Values for $f_+(0)$ determined from the scalar form factor slope using 
the Callan-Treiman relation and  $f_K/f_\pi=1.189(7)$. \label{fig:CTtest} }
\end{figure}

The vector and scalar form factors $f_{+,0}(t)$ are analytic functions in the complex
$t$--plane, except for a cut along the positive real axis, starting at the
first physical threshold $t_{\rm th} = (m_K+m_\pi)^2$, 
where they develop discontinuities. They are real for $t<t_{\rm th}$.
Cauchy's theorem implies that $f_{+,0}(t)$ can be written as 
a dispersive integral along the physical cut where all possible on-shell 
intermediate states contribute to its imaginary part.
A number of subtractions is needed to make the integral convergent.
Particularly appealing is an improved dispersion 
relation recently proposed 
where two subtractions are performed at $t=0$
(where by definition, $\tilde f_0(0)\equiv 1$) and at
the so-called Callan-Treiman point $t_{CT} \equiv (m_K^2-m_\pi^2)$.
Since the Callan-Treiman relation fixes the value of scalar form factor at $t_{CT}$
to the ratio $(f_K/f_\pi)/f_+(0)$,
the dispersive parametrization for the scalar form factor 
allows to transform the available measurements of the scalar form factor 
into a precise information on $(f_K/f_\pi)/f_+(0)$, completely independent of
the lattice estimates. 
Figure \ref{fig:CTtest} shows the values for $f_+(0)$ 
determined from the scalar form factor slope
measurements obtained using a dispersive parametrization and the Callan-Treiman relation, and
$f_K/f_\pi=1.189(7)$. from result on the FF slope using the dispersive 
parameterization
The value of $f_+(0)=0.964(5)$ from UKQCD/RBC is also shown.

\section{The special role of of $\Gamma(K_{e2})/\Gamma(K_{\mu2})$}
\label{ke2}
The ratio $R_K = \Gamma({K_{\mu2}})/\Gamma({K_{e2}})$ can be precisely calculated 
within the Standard Model.
Neglecting radiative corrections, it is given by 
$
R_K^{(0)} = \frac{m_e^2}{m_\mu^2} \: \frac{(m_K^2 - m_e^2)^2}{(m_K^2 - m_\mu^2)^2} 
= 2.569 \times 10^{-5},
$
and reflects the strong helicity suppression of the electron channel.
Radiative corrections have been computed with effective theories,
yielding the final SM prediction
$
R^{\rm SM}_K = R_K^{(0)} ( 1 + \delta R_K^{\rm{rad.corr.}})
= 2.569 \times 10^{-5} \times ( 0.9622 \pm 0.0004 ) =(2.477 \pm 0.001) \times 10^{-5}. 
$
Because of the helicity suppression within then SM, the 
$K_{e2}$  amplitude is a prominent candidate
for possible sizable contributions from physics beyond the SM. Moreover,
when normalizing to the $K_{\mu2}$ rate, we obtain an extremely precise 
prediction of the $K_{e2}$ width within the SM. In order to be visible 
in the $K_{e2}/K_{\mu2}$ ratio, the new physics must  violate lepton 
flavor universality.

Recently it has been pointed out that in a supersymmetric framework 
sizable violations of  lepton universality can be expected
in $K_{l2}$ decays. 
At the tree level, lepton flavor violating terms are forbidden in the MSSM. 
However, these appear at the one-loop level, where an effective 
$H^+ l \nu_\tau$ Yukawa interaction is generated.
The non-SM contribution to $R_K$ can be written as 
$
R_K^{\rm{LFV}} \approx R_K^{\rm{SM}} \left[ 1 + \left( \frac{m_K^4}{M_{H^\pm}^4} \right) \left( \frac{m_\tau^2}{M_e^2} \right) |\Delta_{13}|^2 \tan^6 \beta \right]{\mbox ,}
$ where $\Delta_{13}$, the lepton flavor violating coupling , being generated at the 
loop level, could reach values of $\mathcal{O}(10^{-3})$.
For moderately large $\tan \beta$ values, this contribution may therefore
enhance $R_K$ by up to a few percent.

\begin{figure}[t]
\centering
\includegraphics[width=0.7\linewidth]{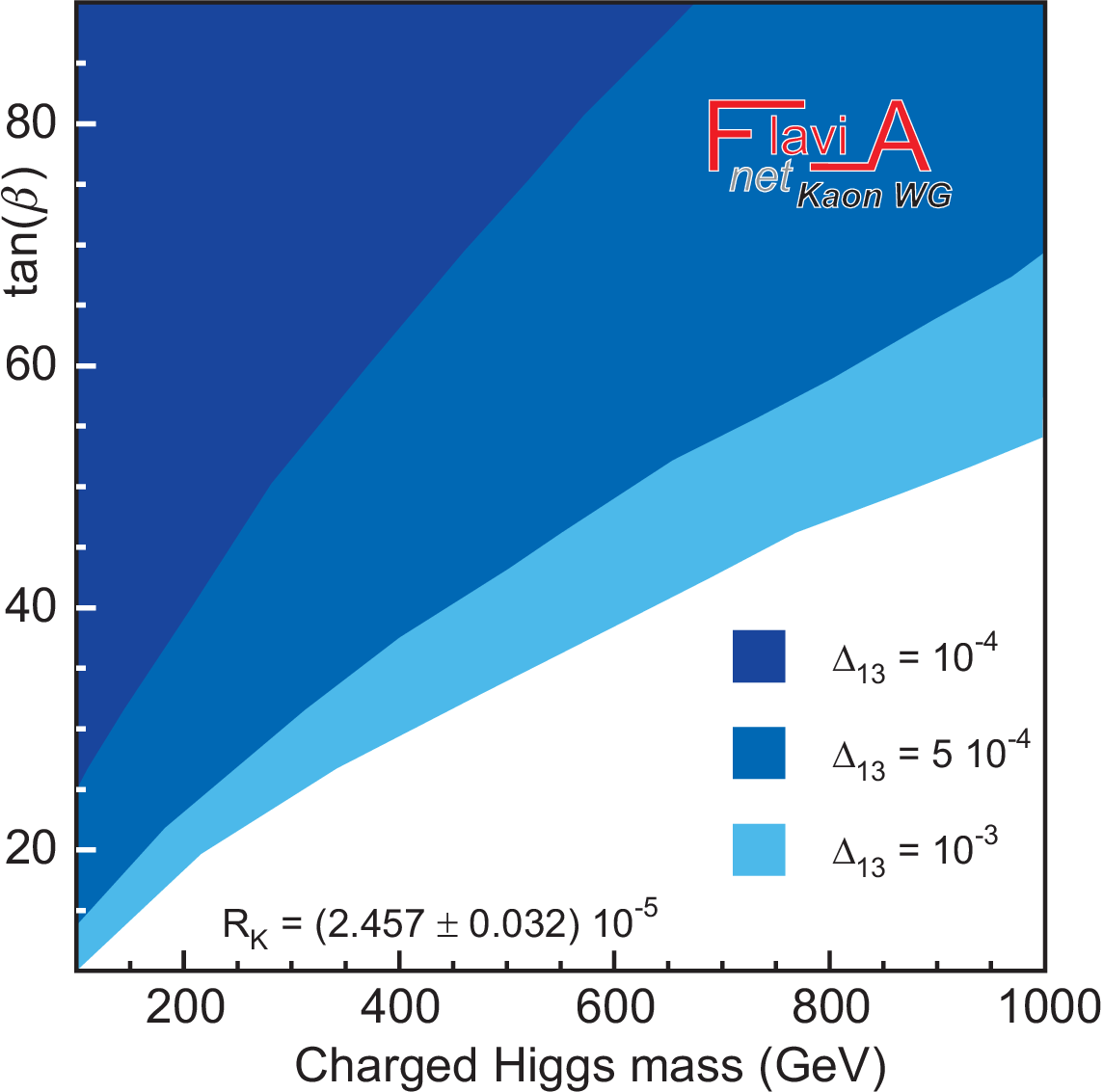}
\caption{Exclusion limits at $95\%$ CL on $\tan \beta$ and the charged
Higgs mass $M_{H^\pm}$ 
from $|V_{us}|_{K\ell2}/|V_{us}|_{K\ell3}$ for different
values of $\Delta_{13}$. }
\label{fig:susylimit}
\end{figure}

Experimental knowledge of $K_{e2}/K_{\mu2}$ has been poor so far.
The current world average
of $R_K = \BR{K_{e2}}/\BR{K_{\mu2}}= (2.45 \pm 0.11) \times 10^{-5}$ dates back to three 
experiments
of the 1970s~\cite{PDG06} and has a precision of about 5\%.
Three new preliminary measurements were reported by NA48/2 and KLOE
 (see~\cite{Flavia2008} for details).
Both, the KLOE and the NA48/2 measurements are inclusive with respect to final state 
radiation  contribution due to bremsstrahlung.
Combining these new results with the current PDG value yields a current world average of
$R_K  = ( 2.457 \pm 0.032 ) \times 10^{-5}$, with a relative error of $1.3\%$,
a factor three more precise than the previous world average.
This value is in very good agreement with the SM expectation
and gives strong constraints 
for $\tan \beta$ and $M_{H^\pm}$, as shown in Fig.~\ref{fig:susylimit}.
For values of $\Delta_{13} \approx 5 \times 10^{-4}$
and  $\tan \beta > 50$ the charged Higgs masses is pushed 
above 1000~GeV/$c^2$ at 95\% CL.

\end{document}